# Performance of ZnSe-based scintillators at low temperatures

S. Galkin[1], I. Rybalka[1,2], L. Sidelnikova[1], A. Voloshinovskii[3], H. Kraus[4], V. Mykhaylyk[5*]

[1]Institute for Scintillation Materials, National Academy of Sciences of Ukraine, 60 Nauky Ave., Kharkiv 61001, Ukraine

[2]V.N. Karazin Kharkiv National University, Crystal Physics Department, 4 Svobody Sq., Kharkiv 61022, Ukraine

[3]Physics Department I. Franko National University of Lviv, 8 Kyryla i Mefodiya Str., 79005, Lviv, Ukraine

[4]University of Oxford, Department of Physics, Denys Wilkinson Building, Keble Road, Oxford, OX1 3RH, UK

[4]Diamond Light Source, Harwell Campus, Didcot, OX11 0DE, UK

**Abstract**

Applications that utilize scintillation detectors at low temperatures are growing in number. Many of these require materials with high light yield and a fast response. Here we report on the low-temperature characterisation of ZnSe doped with Al or Te, respectively. The X-ray luminescence and decay curves were measured over the 77- 295 K temperature range, and alpha particle excitation was used to examine scintillation light output and decay kinetics over the range 9-295 K. A significant improvement of the scintillation characteristics was observed at cooling below 100 K. The scintillation light yield of the crystals increases by a factor about two, and the decay time constant decreases by almost an order of magnitude to 0.3-0.4 μs. These improvements enhance the potential of ZnSe-based crystals for application in cryogenic scintillation detectors of ionising radiation.

*corresponding author e-mail: vmikhai@hotmail.com

**Introduction**

Progress achieved in material development as well as availability of instrumentation for the detection of scintillation at cryogenic temperatures, such as photomultipliers [1], avalanche photodiodes [2] and silicon photon counters [3] made the operation of scintillation detectors in the low-temperature range technically feasible. As intrinsic light yield of scintillation materials often increases with cooling [4] the idea to utilise this effect for the enhancement of scintillator performance has been around for a long time [5] and it is regularly revisited [6]. The implementation of this idea in practice, however, faces a few challenges. One fundamental issue is that the scintillation decay time normally increases with cooling [4] as temperature dependences of intensity and decay rate are controlled by the same processes of thermal deactivation of emission centres [7]. This increase of the decay time constant has a detrimental impact on the scintillator ability to distinguish individual events, leading to pile-up and thus limitations on the overall detector performance. Due to these limitations, applications of inorganic scintillators at low temperature are currently limited to experiments with a low rate of events.

The limitation brought about by an increase of the scintillation time constant with cooling is not observed in a group of semiconductors where the temperature dependence of the emission kinetic is governed by the dynamics of the recombination of excitons and donor-acceptor pairs. This leads to a decrease of the decay time constant with cooling [8], thus enabling very bright and fast scintillation at low temperatures [9], [10]. The highly efficient scintillator ZnSe, doped with isoelectronic impurities, belongs to this group. Introduced into practical radiation detection about 30 years ago [11] the scintillator features a couple of important advantages, such as an emission spectrum matching the sensitivity curve of Si-photodiodes, high radiation hardness, low afterglow as well as not being hygroscopic and of low toxicity [12], [13], [14], [15]. The scintillation decay time strongly depends on the type of impurity, varying roughly between 1 and 100 $\mu$s [16]. The doping also affects the emission spectrum and light yield [14]. Therefore, the choice of impurity, concentration as well as thermal treatment of the material allows modification of the scintillation characteristics over a broad range. This in turn enables customisation of the scintillator for specific application requirements. Because of this, zinc selenide scintillators lent themselves to different areas of application, spanning the detection of ionising radiation, i.e. X-ray spectroscopy [17], [18], dosimetry [19] and imaging [20], [21]. A new area has been added to this list recently: following the low-temperature characterisation of scintillation properties [22], ZnSe attracted attention in rare event searches, specifically experiments looking for neutrinoless double beta decay of $^{82}$Se [23]. Finding these elusive events requires a very sensitive detection method that relies on simultaneous measurements of scintillation and phonon signals in a crystal cooled to a temperature well below 1 K. Scintillation is then used to identify spurious events due to natural radioactivity and to suppress backgrounds [24, 25]. An extensive research programme on the optimisation of the ZnSe quality [26] eventually led to outstanding scientific results [27].

The successful application at cryogenic temperatures and evidence of the decrease of decay time with cooling observed in an earlier study of ZnSe scintillator [22] reignited the idea of low-temperature scintillation detection of ionising radiation. In order to harness the material for such a purpose we decided to revisit this issue and investigate the performance of ZnSe-based scintillators at cooling. In this paper, we studied how key parameters - scintillation light yield, decay kinetics and X-ray luminescence spectra of ZnSe doped with Al and Te change with temperature. We show that cooling the crystals causes enhancement of the light yield and also reduces the effective decay time of ZnSe, both translating into improved performance of the scintillator at low temperatures.

**Experiment**

The crystals were grown from the melt by the vertical Bridgman–Stockbarger method in an inert gas (argon) atmosphere, using high-purity graphite crucibles. In the process of growth, the melt was overheated by 50 K while the inert gas pressure was kept at $5\times10^5$ Pa. The samples for measurements (10x10x1 mm$^3$) where cut and polished from the ingots using diamond tools.

For X-ray luminescence measurements the sample was placed in a liquid nitrogen cryostat, equipped with a temperature control system. Luminescence was excited by a pulsed X-ray tube (pulse duration ~1 ns, pulse repetition frequency $10^5$ Hz) with a W-anticathode (voltage 40 kV, current 1 mA). The X-ray beam entering the cryostat through a beryllium window irradiated the sample at 45º to the incident radiation. The emission was then collected in reflection mode through a quartz window of the cryostat. The luminescence spectra were measured using a monochromator (MDR-12) and photomultiplier module (Hamamatsu H9305). The luminescence spectra were recorded in integrated mode (signal obtained by integration of the entire decay curve) and with time resolution (signal in 100 ns window following the end of the X-ray excitation pulse). The spectra presented here are corrected for the spectral response of the experimental setup. The X-ray luminescence decay curves were measured using the time-correlated photon counting method by capturing the entire emission spectrum of the scintillators. The overall temporal resolution of the setup is 5 ns. Because of relatively long decay time of the scintillators under study in comparison with the duration ox X-ray excitation pulse no deconvolution procedure was applied for fitting of measured decay curves.

Temperature changes of scintillation properties were studied using the multi-photon counting technique described elsewhere [28]. The single crystal sample was attached to a copper holder with the $^{241}$Am source, placed behind the sample inside of a He-flow cryostat. The temperature was monitored using a Si-diode sensor and stabilised by a PID controller. The signal, detected by a multi-alkali photomultiplier model 9124A (Electron Tubes Enterprises), was digitized using a fast ADC with 5 ns sampling interval. This allows resolving individual photons and recording single-photon signals. The measurements were carried out while cooling the crystals to avoid a contribution from thermally released charge carriers to the scintillation event.

**Results and discussion**

According to literature, ZnSe-Al and ZnSe-Te scintillators selected for this study have similar light yield but very different decay kinetics - fast and slow, respectively [14]. The crystals are transparent with light yellow coloration which is due to the fundamental absorption edge of zinc selenide in the visible spectral range at 2.7 eV (see Fig.1). Both materials are efficient commercial scintillators and, from the viewpoint of the main objective of this study, comparison of their scintillation characteristics with temperature is of particular interest.

Under X-ray excitation, the crystals exhibit an intense broad emission band peaking at 600 and 640 nm in ZnSe-Al and ZnSe-Te, respectively (see Fig.2a). Decreasing the temperature to 77 K allows to resolve two bands in the red part of the spectra, at 630 and 550 nm, as well as a new short-wavelength peak as shown in Fig.2b. The emission band peaking at ca. 465 nm is assigned to the recombination of excitonic states near the band edge [29]. The assignment is consistent with the fast nanosecond decay kinetics dominating in this band, as evidenced by time-resolved measurements (Fig.2b). The uneven shape of the 465 nm band, i.e. the sharp drop of intensity at short wavelengths, can be readily explained by strong self-absorption of high-energy photons exiting the crystal volume.

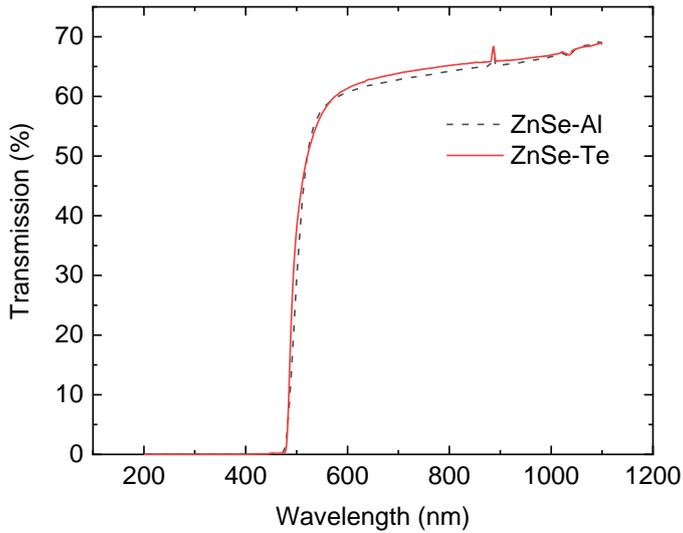

Fig. 1. Transmission spectra of ZnSe-Al and ZnSe-Te crystals at room temperature

The red emission observed in ZnSe-based scintillators is attributed to the radiative recombination of electrons originating from shallow donor centers and different types of defect and impurity complexes formed around ubiquitous zinc vacancies ($V_{Zn}$) and oxygen impurities [16], [30], [31]. Both types of complexes are generic features of the crystal - the result of the technological production process. They are present in the crystals grown by the Bridgeman technique in quantities comparable with other dopants, which impacts the luminescence properties [32]. The low-energy band is believed to be due to acceptor complexes constituting $V_{Zn}$ and impurity ion substituting neighboring host atoms, namely tellurium in ZnSe-Te and aluminum in ZnSe-Al while the high-energy band is traditionally assigned to the recombination of electrons with the oxygen-bearing complex [29], [31]. This assignment is supported by the recent computation study of energy positions of defects in ZnSe [33].

It is reasonable to expect that the concentration of vacancies and types of impurities define the dynamics of changes with temperature of the spectral and kinetic characteristics. At room temperature there is an apparent redistribution of the emission intensity in favor of either the high- or low-energy band, depending on the dopant. The strong luminescence peak at 600 nm observed in ZnSe-Al at room temperature indicates that the recombination of electrons from shallow traps with oxygen-bearing complexes is dominant in this crystal.

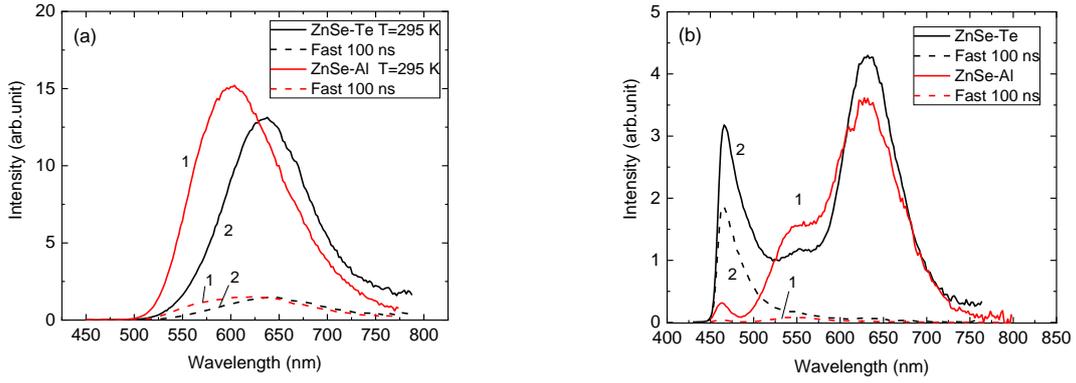

Fig. 2. X-ray luminescence spectra of ZnSe-Al (1) and ZnSe-Te (2) measured at 295 (a) and 77 K (b). Solid line – integrated emission, dotted line – emission detected within the first 100 ns time window.

In ZnSe-Te the main contribution to the emission is due to electron recombination with the acceptor centers $V_{Zn}+Te_{Se}$ resulting in the luminescence shifted towards lower energy. The measurements of scintillation kinetics in the crystals demonstrate how this impacts the luminescence decay. Comparison of the decay curves displayed in Fig. 3 reveals very clearly that at room temperature (or just below) the scintillation kinetics of ZnSe-Al is faster. The rate of radiative recombination in semiconductors depends on the density of donor-acceptor recombination centers: it is faster for higher concentration of the centers [34]. Our observation shows a higher density of emission centers in ZnSe-Al, in line with the previous findings [29].

The decrease of temperature prompts subsequent redistribution of the emission intensity in the X-ray luminescence spectra of the crystals (Fig.2b). In addition to the emerging blue near-band emission, two bands are now well resolved in the red spectral region while the low-energy band is getting more intense in both crystals. Despite of a significant difference in the intensity of the blue bands, in general the spectra of both crystals exhibit qualitative resemblance.

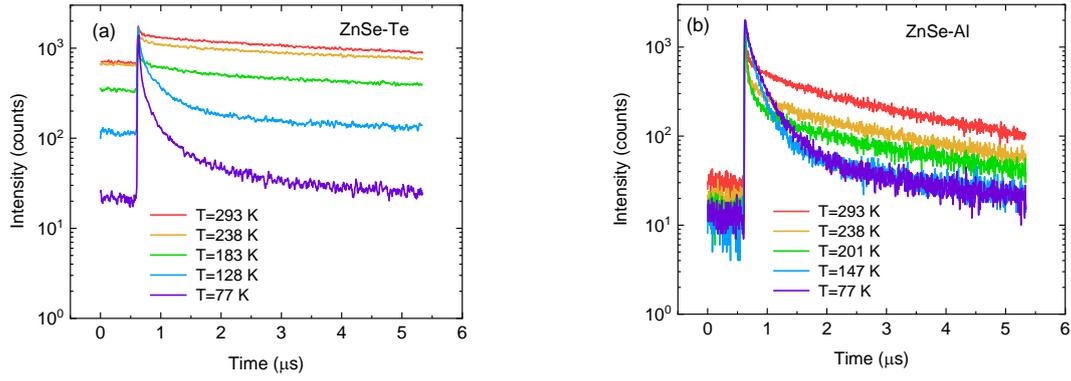

Fig. 3. Luminescence decay curves in ZnSe-Te and ZnSe-Al as a function of temperature, measured at excitation with pulsed X-rays.

The decay curves of the crystals are getting much faster and more alike at low temperatures (see Fig. 3) when compared with those at room temperature. Therefore, we attempted to correlate the features observed in the emission spectra with the scintillation kinetics in the crystals. The measured decay curves exhibit a complex non-exponential shape which is a characteristic feature of the decay kinetics in semiconductors. This is due to a superposition of various recombination processes (donor-acceptor, excitons) exhibiting different types of decay kinetics [8], [34]. In such cases, for the sake of quantitative description, the scintillation decay curves are often approximated by a sum of several exponential functions: $f(t) = y_0 + \sum_i A_i \exp(-t/\tau_i)$, where $y_0$ is background, $A_i$ and $\tau_i$ are amplitudes and decay time constants.

The scintillation decay curves of the crystals were fitted using three exponential functions resulting in R-square better than 0.99. The numerical values derived from the fit are presented in Fig. 4. Despite significant spread of the points, typical for a fit with a large number of variables, one can identify some trends.

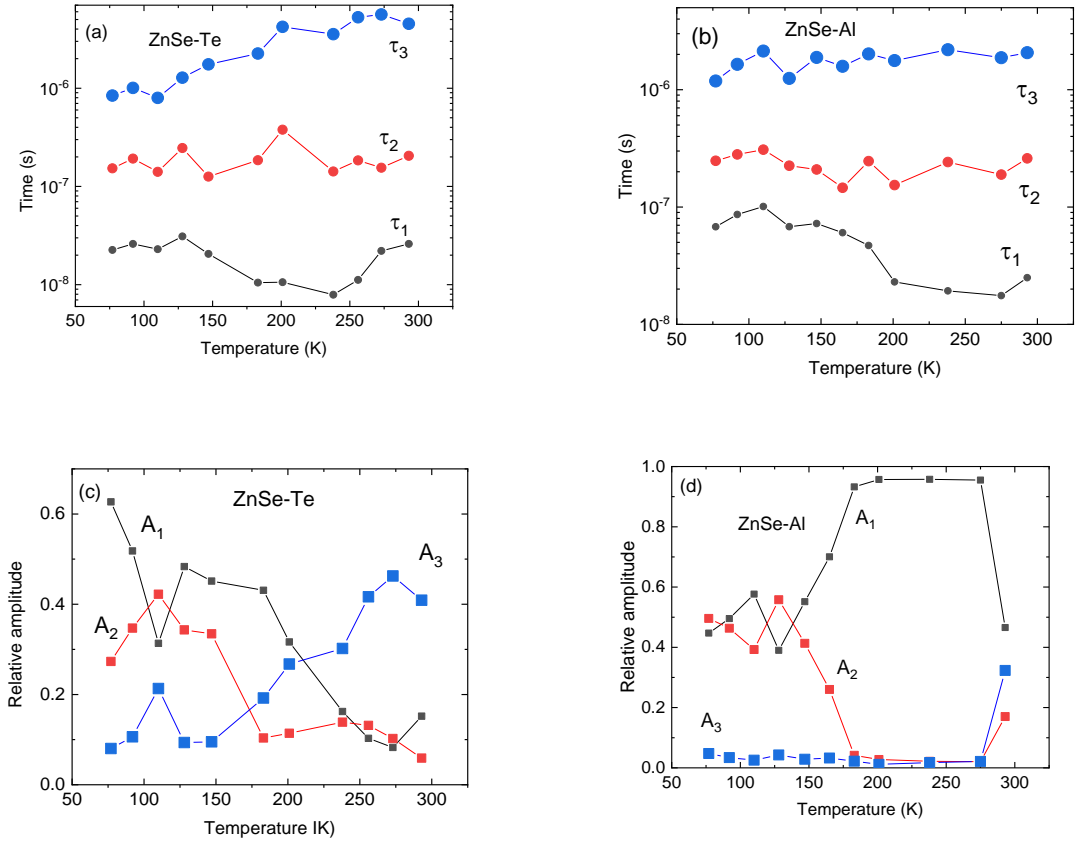

Fig. 4. Parameters of decay curves in ZnSe-Te (a,c) and ZnSe-Al (b,d) as function of temperature. The fractional content of fast, middle and slow amplitudes ($A_1$, $A_2$ and $A_3$) and related decay time constants ($\tau_1$, $\tau_2$ and $\tau_3$) obtained from fitting the decay curves of the crystals with a sum of three exponentials plus background.

The fast component in ZnSe-Te is attributed to the near-band-edge emission emerging in the blue part of the luminescence spectra at cooling (see Fig.2b). This emission increases with decreasing the temperature and essentially this is the main cause of the significant shortening of the decay rate constant. We suggest that at low temperatures more excited carriers can recombine directly from excitonic states without trapping. This is very clearly evidenced by a steady increase of the fractional content of the fast ($A_1$) and middle components ($A_2$) that is accompanied by a decrease of the slow component ($A_3$) with cooling the crystal. Conversely, the decay time constants of ZnSe-Te ($\tau_1$) and ($\tau_2$) exhibit no significant changes with temperature while $\tau_3$ decreases. Altogether, these changes lead to a visible shortening of the scintillation pulse profile at 77 K.

In ZnSe-Al the fast component ($A_1$), dominating at high temperatures, is due to the radiative recombination of emission centres involving oxygen impurities. It drops to ca. 50% below 150 K while the fractional content of the middle component ($A_2$) increases. The fast decay time constant $\tau_1$ also exhibits a fourfold increase at this temperature. The values of $\tau_2$ and $\tau_3$ remain only mildly affected by temperature. The combined effect of these changes is manifested in the alteration of the profile of the luminescence decay pulse: it becomes slower in the initial part while the intensity of the tail (responsible for the long-lasting component) decreases. At 77 K the fractional contribution of the fast component to the overall emission is noticeably smaller in comparison with that of ZnSe-Te that is also seen in the time resolved spectra (Fig.2b).

The results of the fitting were used to derive the temperature dependence of effective decay time constant of the crystals calculated as $\tau_{eff} = \sum A_i \tau_i^2 / \sum A_i \tau_i$ . The plot displayed in Fig.5 reflects the aggregate effect of changes in the parameters that describes the decay curves of the crystals. At cooling from 295 to 77 K the effective decay time constant exhibits about an order of magnitude decrease, i.e. from 1.9 to 0.32 μs in ZnSe-Al and from 4.5 to 0.46 μs in ZnSe-Te.

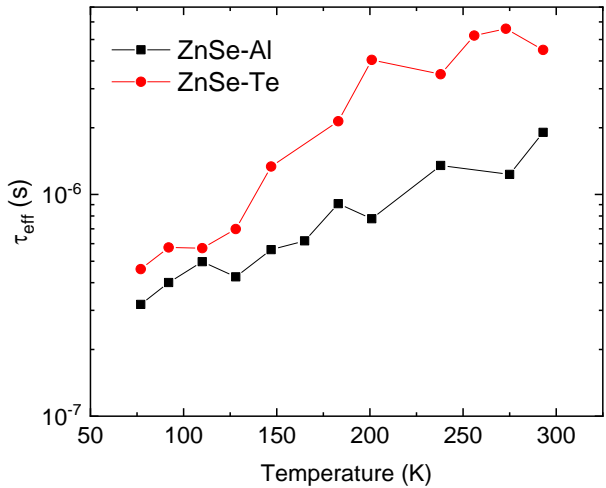

Fig. 5. Effective decay time constant $\tau_{eff}$ of ZnSe-Te and ZnSe-Al as function of temperature.

Further, to characterise the scintillation performance of ZnSe-Te and ZnSe-Al we measured the temperature dependence over the 9-295 K range of the light output and decay curves at excitation with alpha particles. The pulse height spectra of the crystals at different temperatures are shown in Fig. 6. The crystals exhibit very strong scintillation response manifested as clearly resolved peaks, caused by 5.49 MeV α-particles from an $^{241}$Am source depositing their kinetic energy.

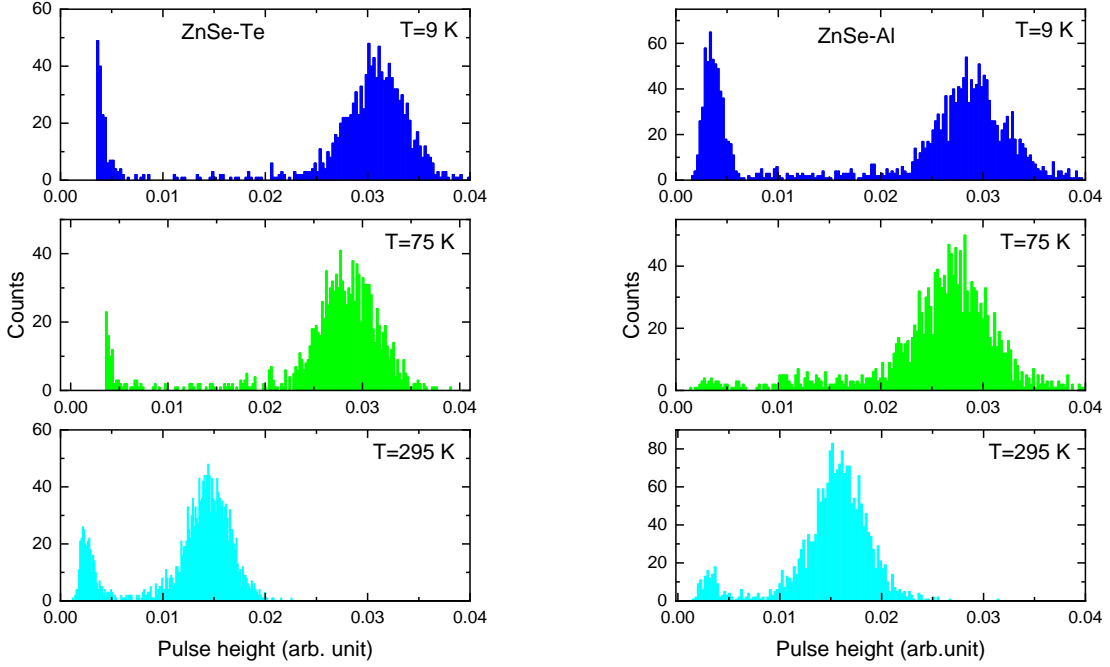

Fig. 6. Pulse height spectra of ZnSe-Te (left) and ZnSe-Al (right) measured at different temperatures at excitation with α-particles from $^{241}$Am decays. The rise of the signal near the origin is due to the detector threshold.

The position of the peak in the pulse height spectra is proportional to the scintillation light yield of a crystal, which enables measurements of the temperature dependence by monitoring the changes in the peak position with cooling [28]. With a decrease of crystal temperature, the peak shifts towards higher amplitudes, evidencing a steady rise of the scintillation light output (see Fig. 7).

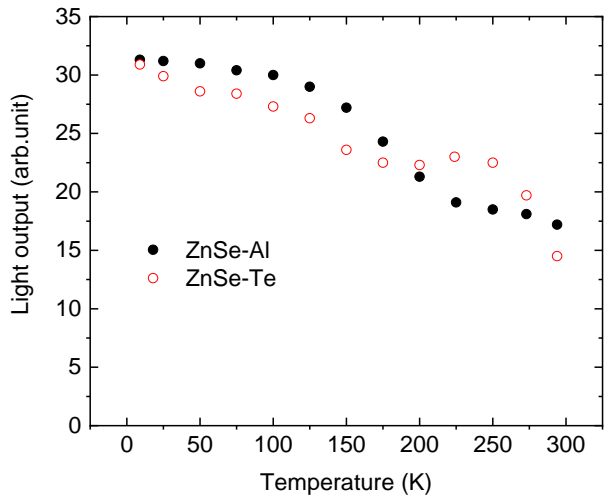

Fig. 7. Scintillation light output of ZnSe-Te and ZnSe-Al as a function of temperature, measured at excitation with α-particles of $^{241}$Am.

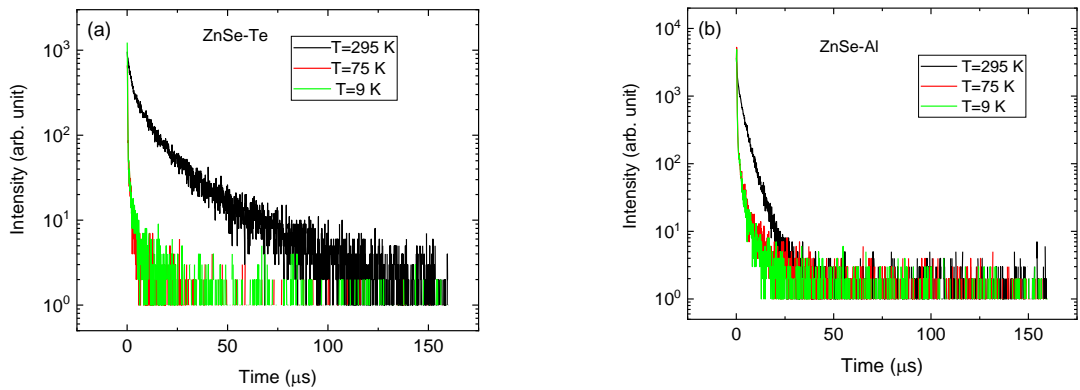

Fig. 8. Scintillation decay curves of ZnSe-Te (a) and ZnSe-Al (b) measured at excitation with α-particles of $^{241}$Am at different temperatures.

The data demonstrate that the scintillation light output at 75 K increases by 80±5 % and 95±5 % in ZnSe-Al and ZnSe-Te, respectively whereas cooling to 9 K causes almost twofold increase of scintillation response of both scintillators in comparison with the room temperature value. Importantly, this enhancement is happening while another key characteristic of the scintillator i.e. response time, is also improving, as was discussed earlier. This is additionally supported by measurements of the scintillation kinetics over a longer time interval. The decay curves shown in Fig. 8 illustrate the effect of significant shortening in the scintillation response of the crystals with cooling, especially in ZnSe-Te, which is consistent with results measured at X-ray excitation. Finally, it is pertinent to remark that cooling to even lower temperatures has no additional visible impact upon the scintillation decay kinetics of the crystals.

**Conclusion**

We carried out measurements of X-ray luminescence decay kinetics and light yield as function of temperature in ZnSe-Al and ZnSe-Te crystal scintillators. The emission of the crystals is attributed to the radiative recombination of electrons captured by shallow traps and acceptor centers, constituting of a Zn vacancy and impurity-ion-substituting neighboring host atoms, i.e. oxygen and tellurium in ZnSe-Te, and oxygen and aluminum in ZnSe-Al. The concentration of the emission centers and the type of impurities determines the distribution of emission intensity in spectra, dynamics of radiative recombination as well as their changes with temperature. Thus, the higher decay rate observed in ZnSe-Al at room temperature is attributed to the higher concentration of recombination centers. A decrease of temperature causes a new high-energy emission band with fast decay to emerge, which is due to recombination of excitonic states near the band edge. The temperature change also causes redistribution of the emission intensity in other parts of the luminescence spectra and an increase of light yield. However, it is the decay kinetic of ZnSe-Te that experiences the major change. Cooling the crystal to below 100 K prompts a redistribution of the intensity in the scintillation pulse in favor of a fast component and a decrease of the slow decay constant, leading to a prominent reduction of the scintillator response time. Moreover, cooling the crystals to this temperature also results in an 80-95% increase of the scintillation light yield. These findings demonstrate that doped ZnSe scintillators cooled to moderately low temperatures just below 100 K exhibit a substantial improvement of performance through an increase of scintillation light yield and a reduction of response time. Taking into consideration that such temperatures are technically easy to achieve, while suitable light detectors are readily available, results of this study pave the way for the development of low-temperature detectors of ionizing radiation using ZnSe-based scintillators.